\documentclass[twocolumn,showpacs,preprintnumbers,amsmath,amssymb,floatfix]{revtex4}
\usepackage{epsf}
 \newcommand{\insertplot}[5]{\begin{figure}
 \hfill\hbox to 0.05in{\vbox to #5in{\vfill
 \inputplot{#1}{#4}{#5}}\hfill}
 \hfill\vspace{-.1in}
 \caption{#2}\label{#3}
 \end{figure}}
 \newcommand{\inputplot}[3]{
 \special{ps: plotfile #1}
}
\begin{document}
\title{Global Charges of Stationary Non-Abelian Black Holes}
\author{
{\bf Burkhard Kleihaus}
}
\affiliation{
{Department of Mathematical Physics, University College, Dublin,
Belfield, Dublin 4, Ireland}
}
\author{
{\bf Jutta Kunz}
}
\affiliation{
{Fachbereich Physik, Universit\"at Oldenburg, 
D-26111 Oldenburg, Germany}
}
\author{
{\bf Francisco Navarro-L\'erida}
}
\affiliation{
{Dept. de F\'{\i}sica Te\'orica II, Ciencias F\'{\i}sicas,
Universidad Complutense de Madrid, E-28040 Madrid, Spain}
}
\date{\today}

\begin{abstract}
We consider stationary axially symmetric
black holes in SU(2) Einstein-Yang-Mills-dilaton theory.
We present a mass formula for these stationary non-Abelian black holes, 
which also holds for Abelian black holes.
The presence of the dilaton field allows for rotating black holes, 
which possess non-trivial electric and magnetic gauge fields, 
but don't carry a non-Abelian charge.
\end{abstract}

\maketitle

{\sl Introduction}
Black holes in Einstein-Maxwell (EM) theory are uniquely characterized
by their global charges:
their mass $M$, their angular momentum $J$, 
their electric charge $Q$, and their magnetic charge $P$
\cite{nohair1,nohair2}.
For EM black holes remarkable relations between their
horizon properties and their global charges hold \cite{nohair2},
such as the Smarr formula \cite{sma},
\begin{equation}
M = 2 TS + 2 \Omega J + \Psi_{\rm el} Q + \Psi_{\rm mag} P
\ , \label{smarr} \end{equation}
where $T$ represents the temperature of the black holes
and $S$ their entropy,
$\Omega$ denotes their horizon angular velocity,
and $\Psi_{\rm el}$ and $\Psi_{\rm mag}$ 
represent their horizon electrostatic and magnetic potential, respectively.

When non-Abelian fields are coupled to gravity, black hole solutions
are no longer uniquely characterized by these global charges
\cite{su2bh,review}.
Thus the EM ``no-hair'' theorem does not readily generalize to theories with
non-Abelian gauge fields coupled to gravity,
and neither does the mass formula, Eq.~(\ref{smarr}) \cite{sudwald,iso1}.

SU(2) Einstein-Yang-Mills (EYM) theory, for instance, possesses sequences of 
static spherically and axially symmetric hairy black hole solutions,
which carry non-Abelian magnetic fields but no non-Abelian charge
\cite{su2bh,kk}; 
it further possesses sequences of
rotating EYM black holes, which carry non-Abelian electric 
and magnetic fields, but only a non-Abelian electric charge \cite{vs,kkrot}.

In many unified theories, including Kaluza-Klein theory
and string theory, a scalar dilaton field arises naturally.
When a dilaton field is coupled to EM theory, 
this has profound consequences for the black hole solutions 
\cite{emd,FZB,Rasheed}.
Although uncharged Einstein-Maxwell-dilaton (EMD) black holes
simply correspond to the EM black holes,
charged EMD black hole solutions possess qualitatively new features.
Charged static EMD black hole solutions, for instance,
exist for arbitrarily small horizon size \cite{emd},
and the surface gravity of `extremal' solutions
depends in an essential way on the dilaton coupling constant $\gamma$.
Extremal charged rotating EMD black holes, known exactly only for 
Kaluza-Klein (KK) theory with $\gamma = \sqrt{3}$ \cite{FZB,Rasheed},
can possess non-zero angular momentum,
while their event horizon has zero angular velocity \cite{Rasheed}.

The known EMD black hole solutions are still uniquely characterized by their
mass, their angular momentum, and their electric and magnetic charge;
and the mass formula Eq.~(\ref{smarr}) holds 
for the KK black hole solutions as well \cite{Rasheed}.

Here we consider stationary black hole solutions
of SU(2) Einstein-Yang-Mills-dilaton (EYMD) theory,
and present a mass formula for these black holes.
After showing, that EMD black holes satisfy the mass formula
\begin{equation}
M = 2 TS + 2 \Omega J + \frac{D}{\gamma}+ 2\Psi_{\rm el} Q 
\ , \label{emdmass} \end{equation}
where the dilaton charge $D$ enters instead of the magnetic charge $P$,
we argue that this mass formula holds 
for all non-perturbatively known black hole solutions
of SU(2) EYMD theory \cite{eymd,kk,long}. 
The mass formula Eq.~(\ref{emdmass}) generalizes the mass formula 
obtained previously for static purely magnetic non-Abelian black holes 
\cite{kk}, $M = 2 TS +D/\gamma $.

We then note that, while similar in many respects
to the rotating EYM black holes \cite{kkrot}, the EYMD black holes
possess new features.
In particular, we show that beyond a certain 
dilaton coupling strength $\gamma$,
the presence of the dilaton allows for a new type of black hole:
a rotating black hole which carries both electric and magnetic 
non-Abelian gauge fields but no non-Abelian charge.

{\sl Abelian mass formula}
Let us first consider the mass formula for EMD black hole solutions
of the action
\begin{equation}
S=\int \left ( \frac{R}{16\pi} + L_M \right ) \sqrt{-g} d^4x
\ \label{actiona} \end{equation}
with matter Lagrangian
\begin{equation}
4 \pi L_M=-\frac{1}{2}\partial_\mu \phi \partial^\mu \phi
 -\frac{1}{4} e^{2 \gamma \phi } F_{\mu\nu} F^{\mu\nu}
\ . \label{lagma} \end{equation}

We start from the general expression for the mass of black holes \cite{wald}
\begin{equation}
M=2TS +2\Omega J_{\rm H} 
- \frac{1}{4 \pi } \int_\Sigma R_0^0 \sqrt{-g} dx d\theta d\varphi
\ , \end{equation}  
and express $R_0^0$ with help of the Einstein equations
and the dilaton equation of motion,
\begin{equation}
R_0^0 =  - \frac{1}{\gamma} \frac{1}{\sqrt{-g}}
 \partial_\mu \left(\sqrt{-g} \partial^\mu \phi\right)
 +  2 e^{2 \gamma \phi}   F_{0\alpha} F^{0\alpha} 
\ . \end{equation} 
Evaluating the integral involving the dilaton d'Alembertian
we obtain the dilaton term, $D/\gamma$, in the mass formula \cite{kk}.

We then replace the horizon angular momentum $J_{\rm H}$
by the global angular momentum $J$ \cite{wald},
\begin{equation}
J=J_{\rm H} + \frac{1}{4 \pi} \int_\Sigma  e^{2 \gamma \phi}
   F_{ \varphi \alpha} F^{0\alpha}  \sqrt{-g} dx d\theta d\varphi
\ , \end{equation} 
and obtain
\begin{eqnarray}
&{\displaystyle M-2TS -2\Omega J - \frac{D}{\gamma} =}
\\
&{\displaystyle- \frac{1}{2\pi} \int_\Sigma  e^{2 \gamma \phi}
  \left( F_{0\alpha}+ \Omega F_{\varphi\alpha} \right) F^{0\alpha} 
 \sqrt{-g} dx d\theta d\varphi}
\ . 
\nonumber 
\end{eqnarray}

To evaluate the remaining integral, we make the replacements
$F_{\alpha 0} = \partial_\alpha A_0$ and
$F_{\alpha \varphi} = \partial_\alpha A_\varphi$,
and employ the gauge field equations of motion.
The mass formula Eq.~(\ref{emdmass}) then holds, provided
\begin{equation}
 2 \Psi_{\rm el} Q =\frac{1}{2\pi}  \int_\Sigma
  \partial_\alpha \left[ ( A_0 + \Omega A_\varphi )
  e^{2 \gamma \phi} F^{0\alpha} \sqrt{-g}  \right] dx d\theta d\varphi
\ . \label{zs} \end{equation}                                                                  
To show Eq.~(\ref{zs}) we choose a gauge, where
the gauge potential vanishes at infinity,
and note that
the electrostatic potential $\Psi_{\rm el}$ 
\begin{equation}
\Psi_{\rm el} =
 \chi^\mu A_\mu = A_0 + \Omega A_\varphi 
\ , \end{equation}
defined with Killing vector
$\chi = \xi +\Omega \eta$ ($\xi=\partial_t$, $\eta=\partial_\varphi$)
is constant at the horizon.
Thus we obtain
\begin{equation}
 2 \Psi_{\rm el} Q =  -  \Psi_{\rm el} \frac{1}{2\pi} \int_{\rm H}
 e^{2 \gamma \phi} ({^*F_{\theta\varphi}})   d\theta d\varphi              
\ , \end{equation}
which holds because the conserved charge
$\tilde Q$ \cite{isodil},
\begin{equation}
\tilde Q = -\frac{1}{4 \pi} \int_{\rm H}
 e^{2 \gamma \phi} ({^*F_{\theta\varphi}})   d\theta d\varphi
\ , \end{equation}
does not depend on the choice of 2-sphere,
i.e.,~$\tilde Q(x_{\rm H})= \tilde Q(\infty) = \tilde Q $,
and $\tilde Q =Q$ for $\phi(\infty)=0$.
When the Smarr formula holds \cite{Rasheed},
Eq.~(\ref{emdmass}) implies $D/\gamma = \Psi_{\rm mag} P - \Psi_{\rm el} Q$.

{\sl Non-Abelian black holes}
We now turn to the non-Abelian black holes of the SU(2) EYMD action
with matter Lagrangian
\begin{equation}
4 \pi L_M=-\frac{1}{2}\partial_\mu \phi \partial^\mu \phi
 - \frac{1}{2} e^{2 \gamma \phi } {\rm Tr} (F_{\mu\nu} F^{\mu\nu})
\ , \label{lagm} \end{equation}
field strength tensor
$
F_{\mu \nu} = 
\partial_\mu A_\nu -\partial_\nu A_\mu + i \left[A_\mu , A_\nu \right] 
$
and gauge field $A_\mu = 1/2 \tau^a A_\mu^a$.

For the metric we choose the
stationary axially symmetric Lewis-Papapetrou metric 
in isotropic coordinates,
\begin{equation}
ds^2 = -fdt^2+\frac{m}{f}\left(dx^2+x^2 d\theta^2\right) 
       +\frac{l}{f} x^2 \sin^2\theta
          \left(d\varphi-\frac{\omega}{x}dt\right)^2   
\ . \label{metric} \end{equation}
For the gauge field we employ the ansatz \cite{kkrot},
\begin{equation}
A_\mu dx^\mu 
  =   \Psi dt +A_\varphi (d\varphi-\frac{\omega}{x} dt) 
+\left(\frac{H_1}{x}dx +(1-H_2)d\theta \right)\frac{\tau_\varphi}{2}
\ , \label{a1} \end{equation}
\begin{equation}
A_\varphi=   -\sin\theta\left[H_3 \frac{\tau_x}{2} 
            +(1-H_4) \frac{\tau_\theta}{2}\right] \ , \ \ 
\Psi =B_1 \frac{\tau_x}{2} + B_2 \frac{\tau_\theta}{2}
\ , \label{a3} \end{equation}
where the symbols $\tau_x$, $\tau_\theta$ and $\tau_\varphi$
denote the dot products of the Cartesian vector of Pauli matrices
with the spherical spatial unit vectors.
With respect to the residual gauge degree of freedom \cite{kk} 
we choose the gauge condition 
$x\partial_x H_1-\partial_\theta H_2 =0$ \cite{kkrot}.
All functions depend only on $x$ and $\theta$.
The above ansatz satisfies the Ricci circularity and Frobenius conditions
\cite{wald}.

The event horizon of stationary black holes resides at a surface
of constant radial coordinate $x=x_{\rm H}$,
and is characterized by the condition $f(x_{\rm H})=0$.
At the horizon we impose the boundary conditions
\cite{kkrot,long}
$f=m=l=0$, $\omega=\omega_{\rm H}=\Omega x_{\rm H}$,
$\partial_x \phi = 0$,
$H_1=0$, 
$\partial_x H_2= \partial_x H_3= \partial_x H_4=0$,
$B_1-\Omega \cos\theta=0$,
$B_2+\Omega \sin\theta=0$,
where $\Omega$ is the horizon angular velocity.

The boundary conditions at infinity,
$f=m=l=1$, $\omega=0$, $\phi=0$,
$H_1 =H_3 =  0$, $H_2=H_4 = \pm 1$, $B_1 = B_2 = 0$,
ensure, that black holes are asymptotically flat
and magnetically neutral.
Axial symmetry and regularity impose
the boundary conditions on the symmetry axis ($\theta=0$),
$\partial_\theta f = \partial_\theta l = 
\partial_\theta m = \partial_\theta \omega = 0$,
$\partial_\theta \phi = 0$,
$H_1 =  H_3 = B_2=0$,
$\partial_\theta H_2 = \partial_\theta H_4 = \partial_\theta B_1=0$,
and agree with
the boundary conditions on the $\theta=\pi/2$-axis,
except for $B_1 = 0$, $\partial_\theta B_2 = 0$.

To show the mass formula Eq.~(\ref{emdmass}) for the non-Abelian black holes,
we need to consider the asymptotic expansion for the metric,
the gauge field and the dilaton functions \cite{kkrot,long}.
The mass $M$, the angular momentum $J=aM$,
the non-Abelian electric charge $Q$ 
(where $|Q|$ is the gauge invariant non-Abelian electric charge of 
Ref.~\cite{iso1}),
and the dilaton charge $D$ are obtained from
the asymptotic expansion via
\begin{equation}
f \rightarrow 1 - \frac{2 M}{x} \ , \ \ \
\omega \rightarrow \frac{2 J}{x^2}
\ , \label{MJ}
\end{equation}
\begin{equation}
\left( \cos \theta B_1 + \sin \theta B_2 \right)
\rightarrow \frac{Q}{x} \ , \ \ 
\
\phi \rightarrow - \frac{D}{x}
\ . \label{QD}
\end{equation}

We further need the horizon area $A$ and the
horizon temperature $T= \kappa_{\rm sg}/2 \pi$ with
surface gravity $\kappa_{\rm sg}$ \cite{wald},
\begin{equation}
\kappa^2_{\rm sg} = -1/4 (\nabla_\mu \chi_\nu)(\nabla^\mu \chi^\nu) 
\ . \label{sgwald} \end{equation}

To derive the mass formula for the non-Abelian black holes 
we first follow the arguments employed in the Abelian case.
With the replacements
$F_{\alpha 0} = D_\alpha A_0$ and
$F_{\alpha \varphi} = D_\alpha (A_\varphi-u)$ 
\cite{radu2},
(where $u=\tau_z/2$, and $D_\alpha = \partial_\alpha +i[A_\alpha,\cdot\ ]$,)
the mass formula holds, provided
(see Eq.~(\ref{zs})),
\begin{eqnarray}
& 2 \Psi_{\rm el} Q = \\
& {\displaystyle \frac{1}{4\pi} \int_\Sigma
 {\rm Tr}  \left[  D_\alpha ( A_0 + \Omega (A_\varphi-u) )
 e^{2 \gamma \phi}
 F^{0\alpha}
 \sqrt{-g}\right]  dx d\theta d\varphi}
\ , \nonumber 
\end{eqnarray}

Since the trace of a commutator vanishes, we 
replace the gauge covariant derivative by the partial derivative,
and again make use of the fact that the electrostatic potential is constant 
at the horizon,
$\chi^\mu A_\mu = \Psi_{\rm H} = \Omega u$, 
thus $\Psi_{\rm el}=\Omega$ \cite{long}.
Hence the integral vanishes at the horizon,
and we are left with the integral at infinity. 
Evaluating this integral with help of the
asymptotic expansion \cite{long} then yields the desired result.
Thus the mass formula Eq.~(\ref{emdmass}) holds 
for the non-Abelian black holes as well \cite{long}.

{\sl Numerical results}
We solve the set of eleven coupled non-linear
elliptic partial differential equations numerically,
subject to the above boundary conditions,
employing compactified dimensionless coordinates,
$\bar x = 1-(x_{\rm H}/x)$ \cite{kkrot}.

Starting with a static spherically symmetric 
SU(2) EYMD black hole 
with horizon radius $x_{\rm H}$,
corresponding to $\omega_{\rm H} =0$,
we choose a small but finite value for $\omega_{\rm H}$.
The resulting rotating black hole then has
non-trivial functions $\omega$, $H_1$, $H_3$, $B_1$, $B_2$.
By varying $\omega_{\rm H}$, while keeping the horizon parameter
$x_{\rm H}$ and the dilaton coupling constant $\gamma$ fixed,
we obtain a set of rotating hairy black holes, analogous in many respects
to the EYM black holes \cite{kkrot}.

In Fig.~1 we display the specific angular momentum $a=J/M$,
the non-Abelian electric charge $Q$, and the relative
dilaton charge $D/\gamma$
for black holes with horizon parameter $x_{\rm H}=0.1$
and dilaton coupling constant $\gamma=1$
as functions of the mass $M$.
As expected \cite{kkrot},
these global charges are close to the corresponding
global charges of the embedded Abelian solutions \cite{foot1} 
with $Q=0$ and $P=1$.

\begin{figure}[h!]
\begin{center}
\epsfysize=6cm
\mbox{\epsffile{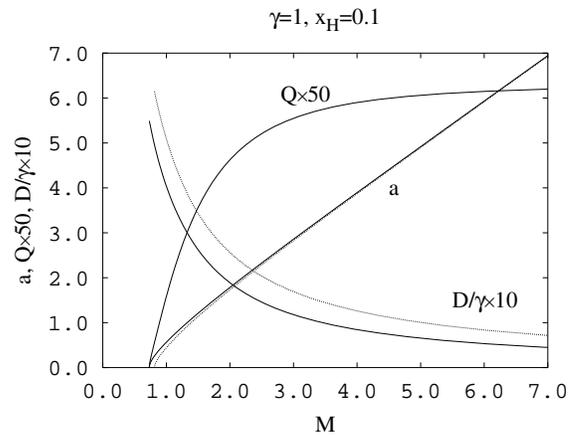}}
\caption{                                                                       
The specific angular momentum $a$,
the non-Abelian electric charge $Q$,
and the relative dilaton charge $D/\gamma$
are shown for the EYMD black hole solutions 
($x_{\rm H}=0.1$, $\gamma=1$) as functions of 
the mass $M$. 
Also shown are the corresponding properties of the embedded Abelian solutions 
with $Q=0$ and $P=1$ (dotted).
}
\end{center}
\end{figure}

In Fig.~2 we exhibit the global charges as functions of
the dilaton coupling constant $\gamma$.
With increasing $\gamma$
the mass $M$ and the relative dilaton charge $D/\gamma$ decrease monotonically,
the specific angular momentum $a$
and the non-Abelian electric charge $Q$ pass extrema \cite{long}.
Interestingly, the non-Abelian electric charge $Q$
can change sign in the presence of the dilaton field.

\begin{figure}[h!]
\begin{center}
\epsfysize=6cm
\mbox{\epsffile{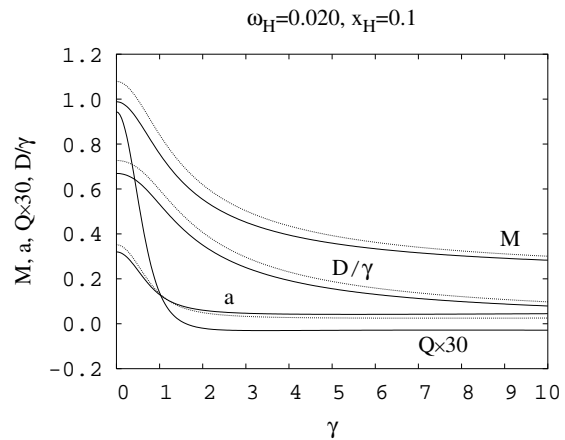}}
\caption{  
The global charges (see Fig.~1) are shown as functions of the
dilaton coupling constant $\gamma$
($x_{\rm H}=0.1$, $\omega_{\rm H}=0.02$).
}
\end{center}
\end{figure}

Thus we observe the surprising feature that the non-Abelian charge $Q$ 
of rotating EYMD black holes can vanish.
Cuts through the parameter space of solutions with vanishing $Q$ 
are exhibited in Fig.~3.
Solutions with $Q=0$ exist only above $\gamma_{\rm min} \approx 1.15$.
These $Q=0$ EYMD black holes represent 
the first black hole solutions, which carry
non-trivial non-Abelian electric and magnetic fields
and no non-Abelian charge \cite{foot2}.
As a consequence, these special solutions do not exhibit the
generic asymptotic non-integer power fall-off of the 
stationary non-Abelian gauge field solutions \cite{kkrot,long}.

\begin{figure}[h!]
\begin{center}
\epsfysize=6cm
 \mbox{\epsffile{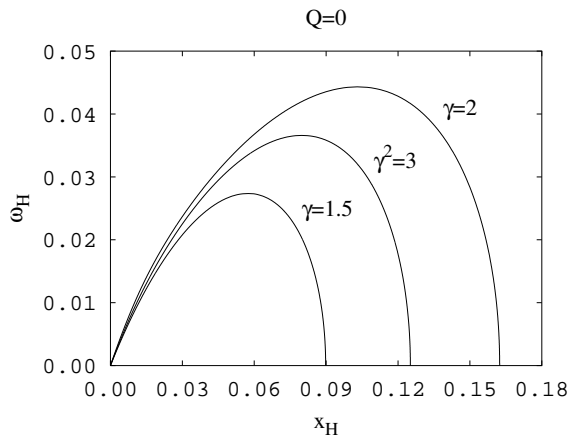}}
\caption{                                                                       
Cuts through the parameter space 
of $Q=0$ black hole solutions
($\gamma=1.5$, $\sqrt{3}$, 2).
}
\end{center}
\end{figure}

Let us now turn to the horizon properties of the EYMD black holes.
In Fig.~4 we show the area parameter $x_\Delta = \sqrt{A/4 \pi}$, 
the temperature $T$,
the deformation of the horizon (as quantified by the ratio of
equatorial and polar circumferences) $L_e/L_p$,
and the Gaussian curvature at the poles $K$.
As for EM black holes \cite{sma},
the Gaussian curvature of the horizon can become negative
and the topology of the horizon is that of a 2-sphere.

 \begin{figure}[b!]
 \vspace{-5mm}
\begin{center}
\epsfysize=6cm
\mbox{\epsffile{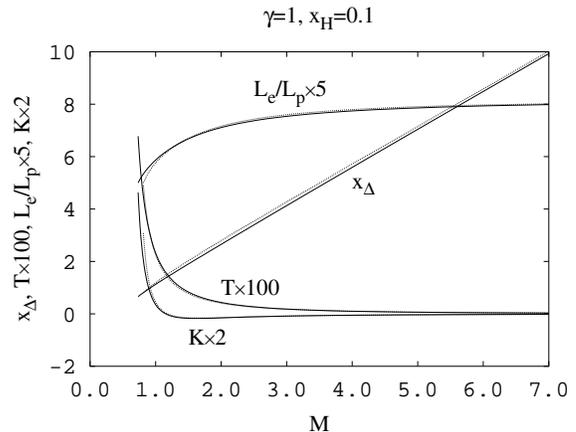}}
\caption{                                                                       
Same as Fig.~1 for the area parameter $x_\Delta$, the temperature $T$,
the deformation of the horizon $L_e/L_p$,
and the Gaussian curvature at the poles $K$.
}
\end{center}
\end{figure}

The numerically constructed stationary axially symmetric EYMD black holes 
satisfy the mass formula, Eq.~(\ref{emdmass}), with an accuracy of $10^{-3}$.
So do the numerically constructed EMD black holes.
EYM black holes are included in the limit $\gamma \rightarrow 0$,
since $D/\gamma$ remains finite.
Further details of the rotating non-Abelian black hole solutions
will be given elsewhere \cite{long}.

{\sl Outlook}
The mass formula holds for the non-perturbatively known
SU(2) EYMD black hole solutions.
However, there may be further black hole solutions in SU(2) EYMD theory,
with different boundary conditions and symmetries.
For such black holes, the mass formula will have to be reconsidered.

Contrary to expectation,
black holes in SU(2) EYMD theory are not uniquely characterized by 
their mass $M$, their angular momentum $J$, 
their non-Abelian electric charge $Q$, and their dilaton charge $D$
\cite{long}. 
Thus a new uniqueness conjecture for non-Abelian black holes will have
to include an additional charge \cite{ash}.

{\sl Acknowledgment}
FNL has been supported in part by a FPI Predoctoral Scholarship from 
Ministerio de Educaci\'on (Spain) and by DGICYT Project PB98-0772.


\begin{thebibliography}{000}
\bibitem{foot2}
 Perturbative studies \cite{vs} suggested that EYM theory possesses
 rotating black hole solutions with non-Abelian electric and magnetic fields
 and no non-Abelian charge, satisfying a different set of boundary conditions.
 Such solutions, however, could not be obtained numerically \cite{radu2,kkrot}.
\end{thebibliography}
\end{document}